\documentclass[useAMS,usenatbib]{mnras}
\usepackage{natbib}
\usepackage{graphicx}

\newcommand{\msun}{\mbox{$\,{\rm M}_\odot$}}

\title[Palomar\,4 on an eccentric orbit]{Direct N-body simulations of globular clusters -- III. Palomar\,4 on an eccentric orbit}

\author[Zonoozi et al.]
{Akram Hasani Zonoozi$^{1}$\thanks{
E-mail:  \mbox{a.hasani@iasbs.ac.ir} (AHZ);
\mbox{haghi@iasbs.ac.ir} (HH)},
Hosein Haghi$^{1}$,  Pavel Kroupa$^{2, 3}$, Andreas H.W. K\"{u}pper$^{4}$\thanks{Hubble Fellow},
\newauthor
Holger Baumgardt$^{5}$ \\
$^{1}$Department of Physics, Institute for Advanced Studies in Basic Sciences (IASBS), PO Box 11365-9161, Zanjan, Iran\\
$^{2}$Helmholtz-Institut f\"ur Strahlen-und Kernphysik (HISKP), Universit\"at Bonn, Nussallee 14-16, D-53115 Bonn, Germany\\
$^{3}$Charles University in Prague, Faculty of Mathematics and Physics, Astronomical Institute, V Hole\v{s}ovi\v{c}k\'ach 2, CZ-180 00 \\
Praha 8, Czech Republic\\
$^{4} $Department of Astronomy, Columbia University, 550 West 120th Street, New York, NY 10027, USA\\
$^{5}$School of Mathematics and Physics, University of Queensland, Brisbane, QLD 4072, Australia\\}

\begin{document}

\date{Accepted .... Received}

\pagerange{\pageref{firstpage}--\pageref{lastpage}} \pubyear{2013}

\maketitle

\label{firstpage}

\maketitle

\begin{abstract}

Palomar\,4 is a low-density globular cluster with a current mass $\approx30000$\msun in the outer halo of the Milky Way with a two-body relaxation time of the order of a Hubble time. Yet, it is strongly mass segregated and contains a stellar mass function depleted of low-mass stars. Pal\,4 was either born this way or it is a result of extraordinary dynamical evolution. Since two-body relaxation cannot explain these signatures alone, enhanced mass loss through tidal shocking may have had a strong influence on Pal\,4. Here, we compute a grid of direct N-body simulations to model Pal\,4 on various eccentric orbits within the Milky Way potential to find likely initial conditions that reproduce its observed mass, half-light radius, stellar MF-slope and line-of-sight velocity dispersion. We find that Pal\,4 is most likely orbiting on an eccentric orbit with an eccentricity of $e\approx 0.9$ and pericentric distance of $R_p\approx5$ kpc. In this scenario, the required 3D half-mass radius at birth is similar to the average sizes of typical GCs ($R_h\approx4-5$\,pc), while its birth mass is about $M_0\approx10^5$\msun. We also find a high degree of primordial mass segregation among the cluster stars, which seems to be necessary in every scenario we considered. Thus, using the tidal effect to constrain the perigalactic distance of the orbit of Pal\,4, we predict that the proper motion of Pal\,4 should be in the range $-0.52\leq\mu_\delta\leq-0.38$ mas\,yr$^{-1}$ and  $-0.30\leq\mu_{\alpha\cos\delta}\leq-0.15$ mas\,yr$^{-1}$.

\end{abstract}


\section{Introduction}

Birth conditions of globular clusters (GCs) such as the initial mass function (IMF), initial binary fraction and the degree of primordial mass segregation are still not well known.  Detailed knowledge of these conditions is necessary to understand their evolution and to get unique insights into star formation in the very early universe. It is therefore essential to specify to what degree the present-day properties of a GC are imprinted by the formation processes and early evolution, and to what extent they are the outcome of long-term dynamical evolution \citep{Vesperini97, Baumgardt03, Heggie03, Leigh15, Marks12, Zonoozi16}.

Two-body relaxation plays an important role in the evolution of characteristic parameters of star clusters (e.g., cluster mass, size, slope of the MF, and  dynamical mass segregation). The majority of Galactic GCs have a present-day median two-body relaxation time, $t_{relax}$, shorter than a Hubble time (Harris 1996, 2010). But there are some unusual GCs like Palomar\,4 and Palomar\,14 (Pal\,4/14, hereafter) with $t_{relax}$ larger than a Hubble time. Therefore, no depletion of low-mass stars and dynamical mass segregation is expected in these clusters as a result of the energy equipartition process.

In Zonoozi et al. (2011, 2014) we performed the first ever full-scale direct N-body simulations of two specific low-mass, extended, remote halo GCs (Pal\,4/14) over their entire life-times (i.e., $11-12$\,Gyr). We compared the results of our simulations to the observed mass, half-light radius, flattened stellar MF and velocity dispersion, which have been determined by Jordi et al. (2009) and Frank et al. (2012). We showed that, for both GCs, dynamical mass segregation alone cannot explain the mass function flattening when starting from a canonical (Kroupa) IMF. Only models with a much larger half-mass radius ($R_h\approx$ 10 pc), and with a flattened IMF \textit{and} a very high degree of primordial mass segregation were able to explain the observations. We concluded that such unusual initial conditions might be obtained by a combination of effects, e.g., through a violent early gas-expulsion phase from an embedded cluster born in a strong tidal field with mass segregation and a canonical IMF (Haghi et al. 2015).

It should be noted that our conclusions were entirely based on the assumption of a \emph{circular orbit} for Pal\,4/14. However, the orbit of Pal\,4 is actually unknown. From observations (Dinescu et al. 1999) and cosmological simulations (Prieto \& Gnedin 2006), we would expect that most of the Galactic GCs are on eccentric orbits.
The mean orbital eccentricity of Galactic GCs with known orbits are around $e\approx 0.5$ \citep{Dinescu07, Dinescu13}.
Therefore it is possible that the orbits of Pal\,4/14 are strongly eccentric. Since star clusters lose more mass during pericentric passages on eccentric orbits, and undergo stronger expansion due to the weaker tidal fields at larger Galactic radii (Baumgardt \& Makino 2003, Webb et al. 2014, K\"{u}pper et al. 2015b), an eccentric cluster orbit might have had an important influence on the evolution of Pal\,4/14, as it could have had a much smaller initial size and significantly higher mass initially.

In fact, energy injection caused by a non-static tidal field leads to the faster ejection of stars, especially during perigalactic passages, where the cluster is tidally shocked. As already found by Webb et al. (2014), for two clusters with the same apogalactic distance, the cluster orbiting in a highly eccentric orbit will undergo a stronger mass loss than the cluster on a circular orbit. Therefore, one could normally expect that for clusters on eccentric orbits the increased tidal stripping leads to severe depletion of low mass stars compared to clusters on circular orbits with the same apogalactic distance. This effect is even more severe in clusters with primordial mass segregation in which the outer part of the clusters are dominated by the low-mass stars, which get preferentially stripped during pericentre passages.

In this work, we perform a comprehensive set of N-body models evolving on eccentric orbits within the Milky-Way potential, by varying the initial conditions and orbits until we find a model that matches the observed properties. Our goal is to determine how tidal shocking can affect the flattening of the stellar MF of Pal\,4 and its mass segregation characteristics.

The paper is organized as follows. In Section 2 the observational  data of Pal\,4 are presented. In Section 3 we describe the set-up of the $N$-body models. This is followed by a presentation of the results of the simulations in Section 4. A discussion of the results and conclusions are presented in Section 5.

\section{Observational constraints on Pal\,4}\label{Sec:Description of the models}

Palomar\,4 has been targeted by several observational programs, which enables us to make detailed comparisons between our numerical simulations and the observational data. In the following section we are going to describe the available constraints and how we use them to set up and analyze our models.

\subsection{Pal\,4's structural characteristics}
We aim to compare the results of our numerical modelling of Pal\,4 with the observational constraints from \citet{Frank12}, who have presented a spectroscopic and photometric study of Pal\,4.  They determined  the cluster's  best-fitting present-day stellar MF-slope inside a projected radius of $r=2.26$\,arcmin ($\approx67$\,pc) in the mass range $0.55\leq m/\msun\leq0.85$ to be $\alpha=1.4\pm 0.25$ which is significantly shallower than a Kroupa (2001) IMF with $\alpha=2.3$ in this mass range. Moreover, Frank et al. found that the slope of the mass function steepens with increasing radius, ranging from $\alpha \leq1$  inside $r\leq1.3~r_{hl}$ to $\alpha \geq2.3$ at the largest observed radii, indicating the presence of mass segregation in Pal\,4.

Another important quantity for comparison is the total observed stellar mass, that is, the mass that has been directly counted from photometric observations (corrected for incompleteness). \citet{Frank12} estimate the total stellar mass inside the projected radius $r \leq2.26$ arcmin, and in the stellar mass range $0.55\leq m/\msun\leq0.85$, to be $5960\pm110\msun$. They do not consider the contributions of blue stragglers and horizontal branch stars in this estimate, which are negligible due to their low numbers.

We also compare the final bound mass of the modelled clusters with the total-mass estimate of \citet{Frank12}, $M_{tot}=29800\pm 800\msun$, which was obtained by extrapolating the measured MF towards lower-mass stars (by assuming $\alpha_{low}=1.3$), and by including the contribution of compact remnants (only white dwarfs, as neutron stars and black holes are expected to escape from the low-density cluster at their formation). The authors also provide a more conservative mass estimate, in which they assume a declining mass function with $\alpha_{low} = -1.0$ in the low-mass range $0.01\leq m/\msun \leq 0.5$, resulting in $M_{tot}=20100\pm 600\msun$ including white dwarfs. We will compare our results to both mass estimates.

By fitting a King (1966)  profile to the surface brightness profile
of Pal 4, Frank et al. found Pal\,4 to have a core radius, a half-light radius, and a tidal radius of $r_c=12.8\pm 1.1$ pc, $r_{hl}=18.4\pm 1.1$ pc, and $r_t=115.5\pm 10.2$ pc, respectively. Assuming mass traces light, the corresponding 3D half-mass radius is about 24 pc. They also estimated Pal\,4 to have an age of $11\pm 1$ Gyr by adopting $[Fe/H]=-1.41\pm 0.17$  for the metallicity from the best fitting isochrone of Dotter et al. (2008). The age estimate is consistent with Pal 4 being part of the young halo population and 1.5--2 Gyr younger than classical, old GCs.  We use this age for our simulations, which start with a stellar population at the zero-age main sequence.

The line-of-sight velocity dispersion of bright stars (with a mass higher than 0.8\msun) measured within the projected radius of $r=50$\,arcsec ($\approx25$\,pc) was determined by Frank et al. (2012) to be $\sigma=0.87\pm0.18$ kms$^{-1}$.

\begin{table*}
\centering
\caption{Summary of the orbital parameters used for the modelling of Pal\,4. The three sets of parameters correspond to the three different types of eccentric orbits we considered.}
\begin{tabular}{ccccccccccccc}
\hline
$R_{gc}$& $X$ & $Y$ & $Z$ &     $V_x$    &    $V_y$     &    $V_z$     & PA  &    $\mu$       &$\mu_{\alpha\cos\delta}$ &$\mu_\delta$    &$R_{p}$& $R_{a}$\\
        &[kpc]&[kpc]&[kpc]&[km\,s$^{-1}$]&[km\,s$^{-1}$]&[km\,s$^{-1}$]&[deg]&[mas\,yr$^{-1}$]&    [mas\,yr$^{-1}$]     &[mas\,yr$^{-1}$]&  [kpc]& [kpc]  \\
\hline
105.5 & $-38.0$& $-12.2$ & 97.7& $-13.4$ & $-3.3$  &  $44.3$ & 205.1 & 0.51 & $-0.22$ & $-0.46$ & 0.5 & 110 \\
-     &        &         &     & $-13.7$ & $-22.7$ & $41.8$ &        & 0.55 & $-0.23$ & $-0.50$ & 5   & 112\\
-     &        &         &     & $-13.9$ & $-36.2$ & $40.1$ &        & 0.58 & $-0.24$ & $-0.52$ & 7   & 115\\
-     &        &         &     & $-14.0$ & $-44.4$ & $39.0$ &        & 0.60 & $-0.25$ & $-0.54$ & 10  & 114\\
\hline
\end{tabular}
\label{tab:orbits}
\end{table*}

\subsection{Assumptions on Pal\,4's orbit}
Palomar\, 4 is located at Galactic coordinates $l= 202.31$\,deg and $b=71.80$\,deg \citep{Harris96}. Using HST/WFPC2 photometry, \citet{Frank12} determined Pal\,4's distance to be $102.8\pm2.4$\,kpc, and with Keck/HIRES data, they measured a radial velocity of $72.55\pm0.22$\,km\,s$^{-1}$. Given Pal\,4's position on the sky, this large receding radial velocity indicates that the cluster may be on an eccentric orbit and moves towards apocenter. That is, put into the Galactic rest frame and assuming standard values for the Sun's position ($R_\odot=8.3$\,kpc) and motion ($U = 11.1$\,km\,s$^{-1}$, $V=254.3$\,km\,s$^{-1}$, $W = 7.25$\,km\,s$^{-1}$) within the Galaxy \citep{Schoenrich12, Kupper15a}, at least 46\,km\,s$^{-1}$ of its heliocentric radial velocity is directed radially outwards.

However, the proper motion of Pal\,4 is unknown, which gives us two free orbital parameter to explore.  Minimising its angular momentum with the coordinate converter tool CoCo\footnote{https://github.com/ahwkuepper/coco}, we find that a position angle of $PA= 205.1$\,deg (measured with respect to Galactic North) and a proper motion of $\mu=0.51$\,mas\,yr$^{-1}$ yields the most eccentric orbit Pal\,4 could be on. For the rest of this paper, we will use this position angle, and only vary the magnitude of the proper motion vector, $\mu$. In this way we study the most extreme orbit for Pal\,4 and orbits similar to that but with slightly smaller eccentricities. A summary of the characteristics of the four orbits we study can be found in Tab.~\ref{tab:orbits}.

\begin{figure}
\begin{center}
\includegraphics[width=85mm]{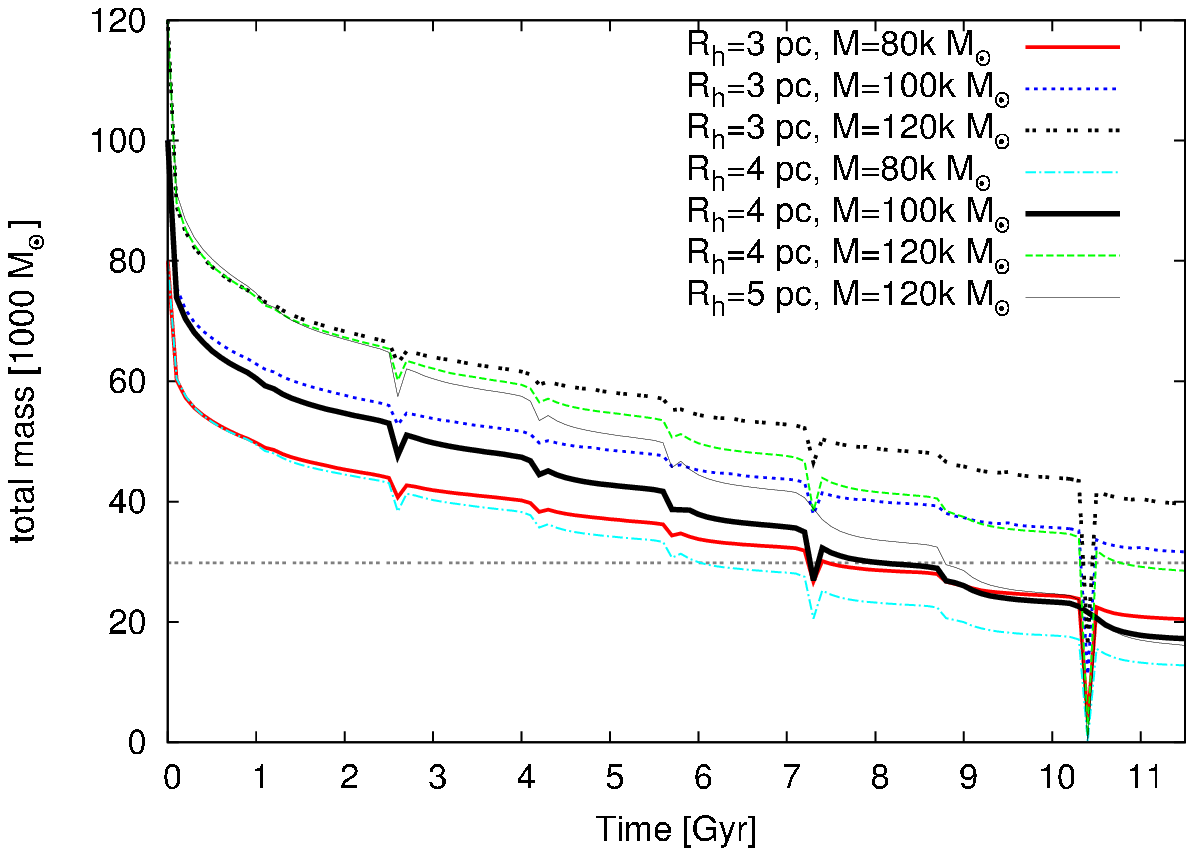}
\includegraphics[width=85mm]{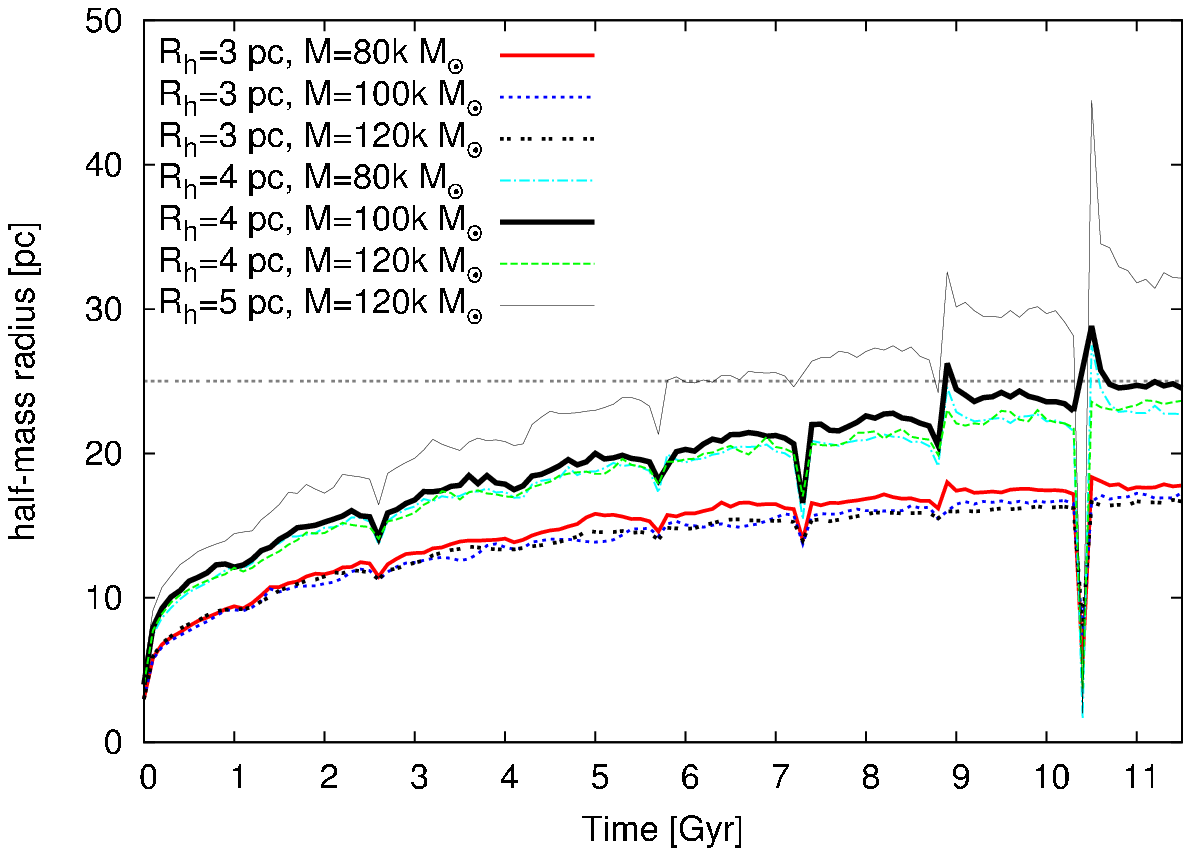}
\caption{The evolution of total mass (top panel) and half mass radius (bottom panel) with time for simulated clusters moving on an eccentric orbit with $R_p=5$ kpc with parameters listed in Table~\ref{tab_regular}. Horizontal dotted lines show Pal\,4's estimated total mass and half-mass radius. Because of stellar evolution, clusters experience a rapid expansion and mass-loss within the first 100 Myr of evolution. Clusters with smaller initial masses and larger initial radii are affected more strongly by the gravitational field of the galaxy. }
\label{massloss}
\end{center}
\end{figure}

\begin{table*}
\centering
\caption{Details of all simulated star clusters starting with different initial masses, half-mass radii and orbital eccentricities. Column 1 gives the model name, in which the first number after 'M' denotes the initial mass in units of 1000\msun  and the second number denotes the initial half-mass radius in parsec. Columns 2 and 3 give the projected half-light radius, $r_{hl}$, and the 3-D half-mass radius,$R_{h}$, of the simulated star clusters after 11 Gyr of evolution. Column 4 gives the present-day (at an age of 11 Gyr) total mass of the modelled clusters in the stellar mass range $0.55\leq m/\msun\leq 0.85$. Column 5 is the present-day total mass of the clusters including low-mass stars and compact remnants. The present-day slope of the mass function, $\alpha_{tot}$, for stars with masses between 0.55 and 0.85\msun inside the projected radius $r=2.26$arcmin, corresponding to $\approx$ 67 pc, is presented in column 6. The model line-of-sight velocity dispersion of bright stars is presented in the last column. Compared to the observational values given at the bottom, only the model number 7 starting with $M=100000\msun$ and $R_h=4$\,pc with a pericentric passage of $R_p=5$\,kpc constitutes an acceptable fit to Pal\,4 after 11 Gyr of evolution and all present-day properties of this model are in good agreement with the observed values. Typical errors of the numerical values given in brackets in the header are calculated by repeating the run for one arbitrary model.}

\begin{tabular}{cccccccc}

\hline (1)&    (2)   &    (3)    &      (4)       &     (5)       &    (6)    &        (7)       \\
\hline
Model   &  $r_{hl}(pc)$  & $R_{h}(pc)$   & $M_{obs}(\msun)$ &$M_{tot}(\msun)$&  $\alpha$ & $\sigma_{los}(km/s)$   \\
        & ($\pm$2.1) &($\pm$2.8) &($\pm$215)      & ($\pm$2500)   &($\pm$0.13)&               ($\pm$0.07)     \\
 \hline
 $R_{p}=5$ kpc &&&&&&&\\
 \hline
 (1)~M120R5 &     22.6   &    31.8   &      5855      &     16976     &    1.41   &          0.63    \\
 (2)~M120R4.5 &     17.7  &    26.3   &      7792      &     22653     &    1.63    &            0.78        \\
 (3)~M120R4 &     14.3   &    23.3   &      9345      &     29158     &    1.87   &            0.92         \\
  (4)~M120R3 &     9.8    &    16.6   &      11310     &     40460     &    2.01  &            1.21         \\
 (5)~M100R5 &     25.3   &    36.8   &      2535      &     7133      &    1.20   &            0.41         \\
 (6)~M100R4.5 &     20.7   &    30.7  &      4110      &     11644     &    1.44   &            0.53       \\
 \textbf{(7*)~M100R4} &     16.3   &    24.7   & 6190  &     18149     &    1.56   &           0.70        \\
 (8)~M100R3 &     10.5   &    17.2   &      9340      &     32361     &    2.10   &            1.07         \\
 (9)~M80R5  &     26.3   &    49.2   &      580       &     1700      &    0.39   &            0.17         \\
 (10)~M80R4  &     14.5   &    22.7   &      4795      &     13229     &    1.63   &           0.61         \\
 (11)~M80R3  &     10.5   &    17.6   &      6630      &     20874     &    1.81   &           0.84         \\
  \hline
 $R_{p}=7$ kpc &&&&&&&\\
 \hline
 (12)~M100R6.5 &     32.5   &    47.7  &      3475      &     10116 &    1.36   &                 0.44        \\
 (13)~M100R6 &     23.0   &    35.8   &      6330      &     18254      &    1.62   &            0.61         \\
 (14)~M100R5 &     17.5   &    29.0   &      8457      &     26235     &    1.87   &             0.78         \\
 (15)~M80R6 &     24.2   &    36.7   &      3921      &    11027  &    1.57   &                    0.47        \\
 (16)~M80R5.5 &     22.3   &   33.7   &      4607      &     12720  &    1.67   &                 0.52     \\
 (17)~M80R5 &     16.7   &    27.9   &      6562      &     19800     &    1.91   &              0.70        \\
 (18)~M70R5.5 &   22.3   &    34.1 &      3379      &     9180   &    1.58        &               0.43    \\
 (19)~M70R5  &   18.5   &    30.8 &      4296      &     11896  &    1.48        &          0.52      \\

 \hline
  $R_{p}=10$ kpc &&&&&&&\\
 \hline
 (20)~~M80R6.5 &   30.5   &    42.0  &      4013      &   11555    &    1.54        &          0.47        \\
 (21)~~M80R6.1 &   22.0   &    35.4  &      6044     &    17757   &    1.81        &           0.60       \\
 (22)~~M80R6 &   20.0   &    33.7  &      6384      &    18916     &    1.88        &           0.65       \\
 (23)~~M80R5 &   15.0   &    26.6  &      7347      &    23303     &    2.01         &           0.77       \\
 (24)~~M70R6 &   24.5   &    35.0  &      4537      &   13065      &    1.75        &           0.53       \\
 (25)~~M70R5 &   18.3   &    29.1  &      5883      &   18008      &    1.84        &           0.65       \\
 (26)~~M65R5 &   20.5   &    29.3  &      5457      &   15500      &    2.06        &           0.62       \\

 \hline
Observations ($\alpha_{low} = 1.3$)&18.4$\pm$1.1&       & 5960$\pm$110   & 29800$\pm$800 & 1.4$\pm$0.25 &             0.87$\pm$0.18     \\
Observations ($\alpha_{low} = -1.0$)                     &                      &       &                            & 20100$\pm$600 &                        &                                           \\

       \hline
\end{tabular}
\label{tab_regular}
\end{table*}

\begin{figure}
\includegraphics[width=85mm]{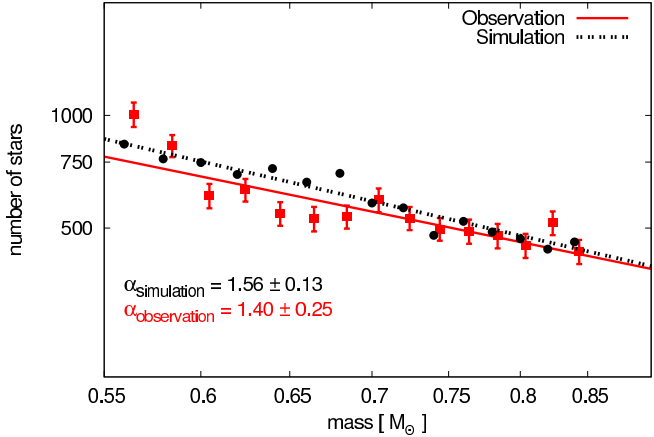}
\caption{Global MF in the mass range 0.55-0.85\msun for model 7, 'M100R4', which started with $M=100000$\msun and $R_h=4$\,pc with a pericentric radius of $R_p=5$\,kpc.The mass function at the start of the simulation was chosen to be a canonical IMF, with $\alpha=2.3$ for stars more massive than 0.5\msun. The red solid line together with the red data points depicts the observed present-day mass function of Pal\,4. The black, dotted line together with
the black data points with a slope of $\alpha=1.56\pm 0.13$ shows the clusters mass function after an evolution of 11 Gyr. Both are in good agreement with the observed value within the uncertainties. The error of the slope of the mass function is derived from least square fitting. Hence, two-body relaxation together with tidal stripping of low-mass stars is able to deplete the mass function sufficiently of low-mass stars in order to reproduce the
observations. } \label{mf}
\end{figure}
\begin{figure}
\includegraphics[width=85mm]{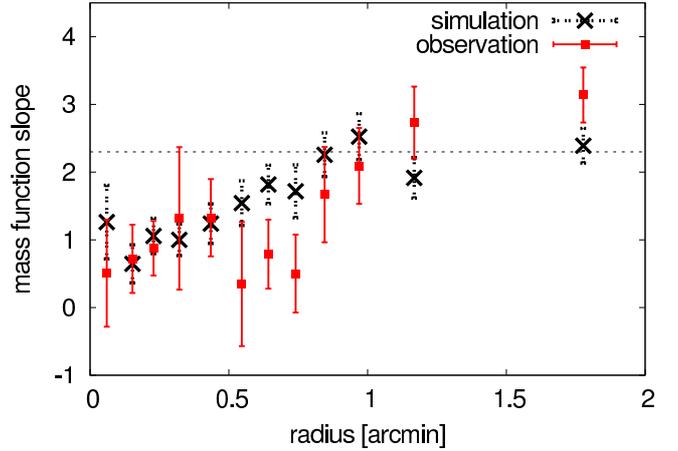}
\caption{ The present-day MF-slope, $\alpha$, derived in different radial
bins for model 'M100R4'. The red squares are the observed values of Pal\,4 taken from Frank et
al. (2012), and the black dots represents the result of the simulation after an
evolution of 11 Gyr. The flattened slope within the inner parts with respect to the
slope in the outer parts indicates that dynamical mass segregation has happened in
the cluster.
 } \label{mfr}
\end{figure}

\begin{figure}
\includegraphics[width=85mm]{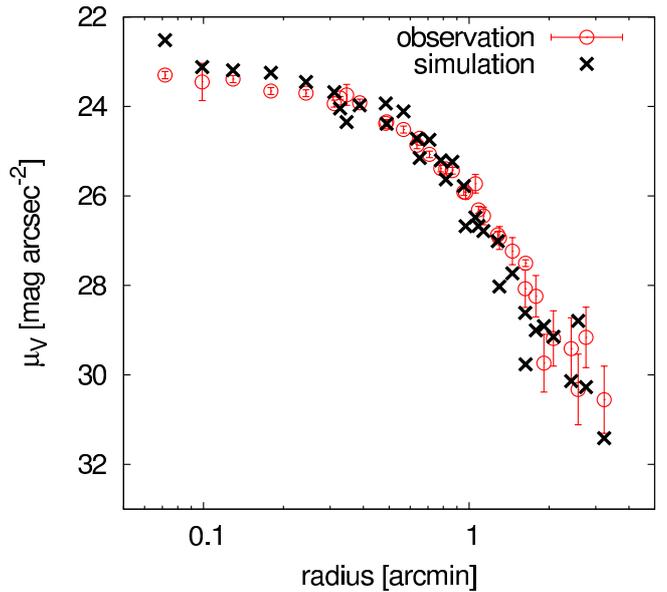}
\caption{The surface brightness profile of Pal 4 taken from Frank et al. (2012, red circles) and the best-fitting model ('M100R4', black crosses). The surface brightness profile of the best fit cluster model is in good agreement with the data.} \label{SB}
\end{figure}

\begin{figure}
\includegraphics[width=85mm]{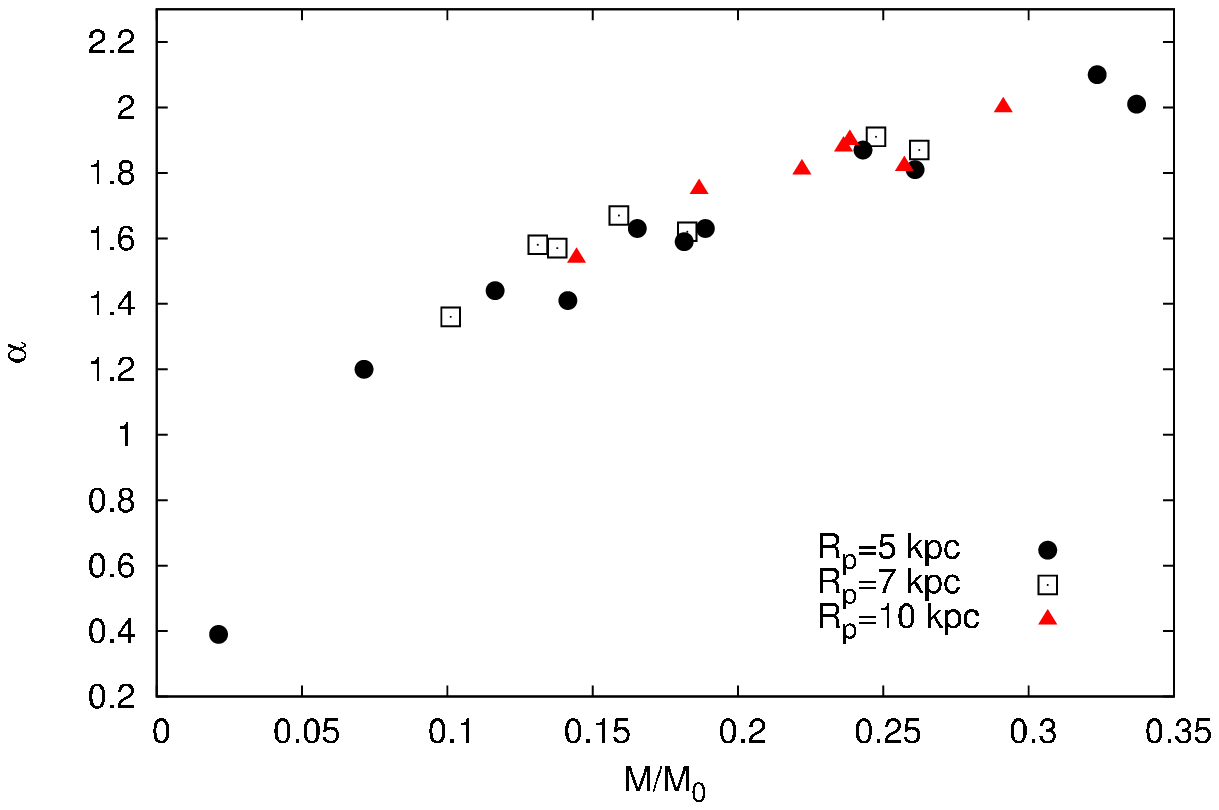}
\caption{ The present-day slope of the stellar mass function at 11 Gyr versus the fraction of mass remaining for each modeled cluster listed in Table 2. The MF slope is a strong function of the mass loss and nearly independent of orbital eccentricity, initial cluster size and mass. Models with $R_p=$5, 7, and 10 kpc are indicated by circles, squares and triangles, respectively. } \label{m-alpha}
\end{figure}

\section{N-body models and choice of initial conditions}\label{Sec:Description of the models}

In this section, we discuss our choices for initial mass, half-mass radius, stellar mass function, and orbital parameters of the modelled star clusters.

We used Sverre Aarseth's freely available, collisional $N$-body code \textsc{Nbody6} \citep{Aarseth03, Nitadori12} on the GPU computers of the University of Queensland to model the dynamical evolution of Pal~4 over its entire lifetime. The code includes a comprehensive treatment of single and binary stellar evolution from the zero-age main sequence (ZAMS) through remnant phases by using the \textsc{SSE/BSE} routines and analytical fitting functions developed by ~\citet{Hurley00} and \citet{Hurley02}. Because of the low escape velocity of Pal~4 over most of its lifetime, we assume a zero retention fraction of neutron stars/black holes (which form during the simulation). 

The initial configuration for all model clusters was set up using the publicly available code \textsc{McLuster}\footnote{\tt  https://github.com/ahwkuepper/mcluster} \citep{Kupper11}. We chose a Plummer model \citep{Plummer11} in virial equilibrium as the initial density profile for our models. We ran simulations with different numbers of stars in the range $N=120000$ to $210000$, corresponding to initial cluster masses of $M_{i}=70000$ and $120000\msun$, respectively. We then followed the evolution of the clusters for 11 Gyr, during which they typically lost about 70-85\% of their mass into the Galactic halo.

The size scale of the Plummer models was set by the initial half-mass radius. Compared to the models evolving on a circular orbit (see Zonoozi et al. 2014), we chose a smaller initial radius (by a factor of $\leq0.5$) and a larger initial mass (by a factor of $\geq2$). This is because of  the enhanced mass-loss rate driven by tidal shocks as the cluster passes through pericentre (in addition to the mass loss by stellar evolution and from dynamical evolution). Therefore we used initial 3D half-mass radii in the range of 3 to 6 pc in order to reach 3D half-mass radii of about 24 pc after 11 Gyr.

We chose the stellar masses according to a canonical IMF \citep{Kroupa01} between the mass limits of 0.08 and 100\msun.  All stars are assumed to be on the zero-age main sequence when the simulation begins at $t=0$. We also assume that any residual gas from the star formation process has been removed. Our model is thus a post-gas-expulsion cluster probably about four times larger than when it was in its embedded phase (Brinkmann et al. 2016).

The code \textsc{McLuster} allows to initialize  any degree of primordial mass segregation (hereafter, $S$) to all available density profiles using the routine described in \citet{Baumgardt08a}. This routine allows to maintain the desired mass density profile when increasing the degree of mass segregation while also making sure that the cluster is in virial equilibrium. For a fully segregated cluster, the most massive star occupies the orbit with the lowest energy, the second most massive star occupies the orbit with the second lowest energy and so forth.  $S=0$ means no segregation and $S=1$ refers to full segregation. For all our initial models we assume $S=0.9$ except one model  which we set to $S=0$ for comparison.

We do not include primordial binaries in our simulated clusters, although binaries can form via three-body interactions throughout the simulations. Having a significant fraction of primordial binaries is computationally too demanding for such clusters as we model here. Moreover, the dynamical effects from primordial binaries are not expected to be significant for such an extended star cluster as Pal\,4 \citep{Kroupa95}.

In all our models, tidal effects of an analytic galactic background potential (consisting of a bulge, a disc and a dark halo component) are included.
The bulge is modeled as a central point-mass:
\begin{equation}
 \Phi_b \left(r\right) = -\frac{G M_b}{r},
\end{equation}
where $M_b$ is the mass of the bulge component. The disk component of the galaxy is modeled following the prescriptions of \citet{Miyamoto75}:
\begin{equation}
\Phi_d \left(x,y,z\right) = -\frac{G M_d}{\sqrt{x^2+y^2+\left(a + \sqrt{z^2+b^2}\right)^2}},
\end{equation}
where $a$ is the disk scale length, $b$ is the disk scale height, and $M_d$ is the total mass of the disk component. We used values of $a= 4$\,kpc and $b=0.5$\,kpc, while for the disk and bulge masses we adopted $M_d = 5\times 10^{10}\msun$ and $M_b = 1.5 \times 10^{10}\msun$, respectively, as suggested by \citet{Xue08} for a Milky Way-like potential.  We use a two-parameter logarithmic DM halo potential of the form

\begin{eqnarray}
\Phi_h (R)= \frac{v_0^2}{2} ~ \ln(R^2+R_c^2).
\end{eqnarray}

Here, $R=x^2+y^2+z^2$ is the distance from the galactic centre at any given time.  The constant $R_c$ is chosen such that the combined potential of the three components yields a circular velocity of $v_0=220$ km/s in the disk plane at a distance of 8.3 kpc from the galactic centre. 

The eccentric orbits of the clusters are chosen to have pericentric distances in the range $R_p\leq10$\,kpc by scaling the proper motion vector magnitude $\mu$ (see Tab.~\ref{tab:orbits}). We use values of $\mu$ in the range of $0.51-0.60$\,mas\,yr$^{-1}$, corresponding to 3D velocities of  $46.6-60.8$\,km\,s$^{-1}$ in the Galactic rest frame. Within the Allen \& Santillan galactic potential, these velocities result in orbits with orbital eccentricities of $e\geq 0.8$.

Comparing the results of this new set of models with our previous investigation \citep{Zonoozi14} shows that Pal~4 on an elliptical orbit with an eccentricity of 0.9 and apogalacticon of about $110$ kpc will have a larger overall mass loss rate by a factor of 3 than the same cluster on a circular orbit at that distance. However, we find that the orbit with the highest possible eccentricity ($\mu=0.51$\,mas\,yr$^{-1}$) brings the clusters to within less than 1\,kpc of the Galactic centre. Such an extreme orbit causes episodes of dramatic mass loss and leads to a rapid dissolution.   We therefore restrict our investigation to orbits with pericentres $R_p\geq 5$\,kpc, but keep the minimum angular momentum orbit in Tab.~\ref{tab:orbits} for completeness.  We evolve three sets of models with different pericentric distances of $R_p$=5, 7, and 10 kpc.  Clusters with different $R_p$ experience different degrees of tidal stripping, and hence can have different loss rates of  low-mass stars  from the outer regions of the clusters.

The choices for the initial conditions of all models with model name based on initial cluster mass and 3D half-mass radius, and the key results of the simulated  clusters are summarized in Table~\ref{tab_regular}.

\begin{figure*}
\includegraphics[width=150 mm]{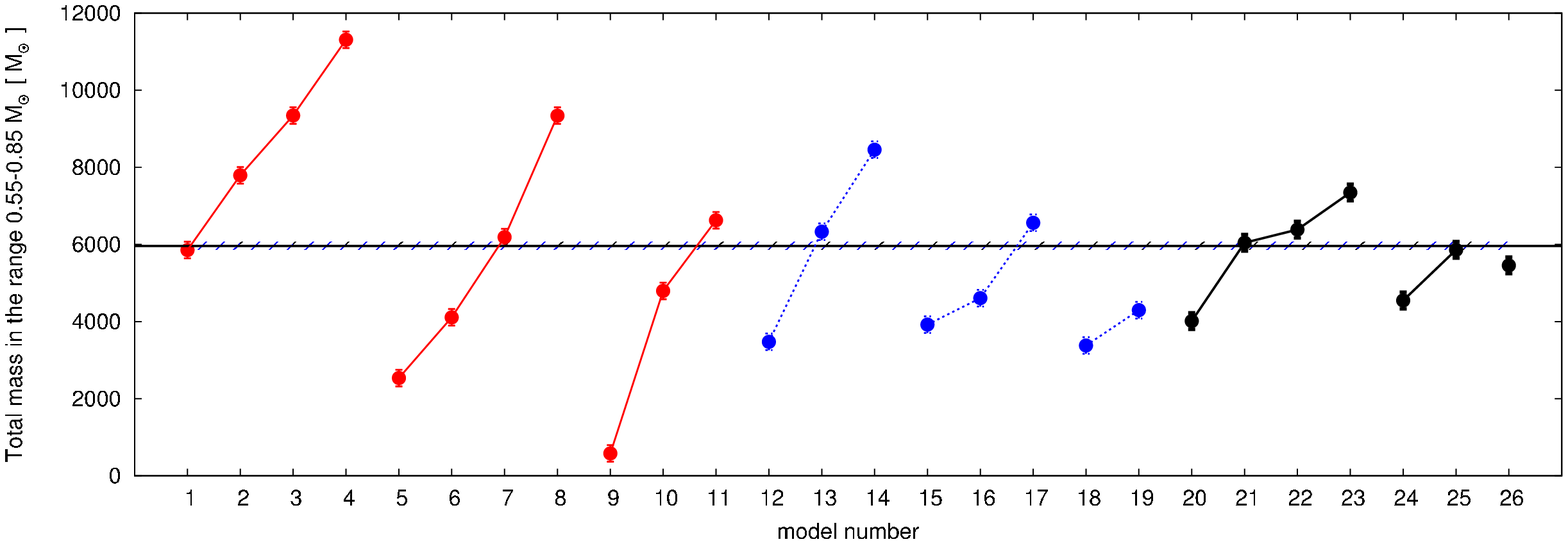}
\includegraphics[width=150 mm]{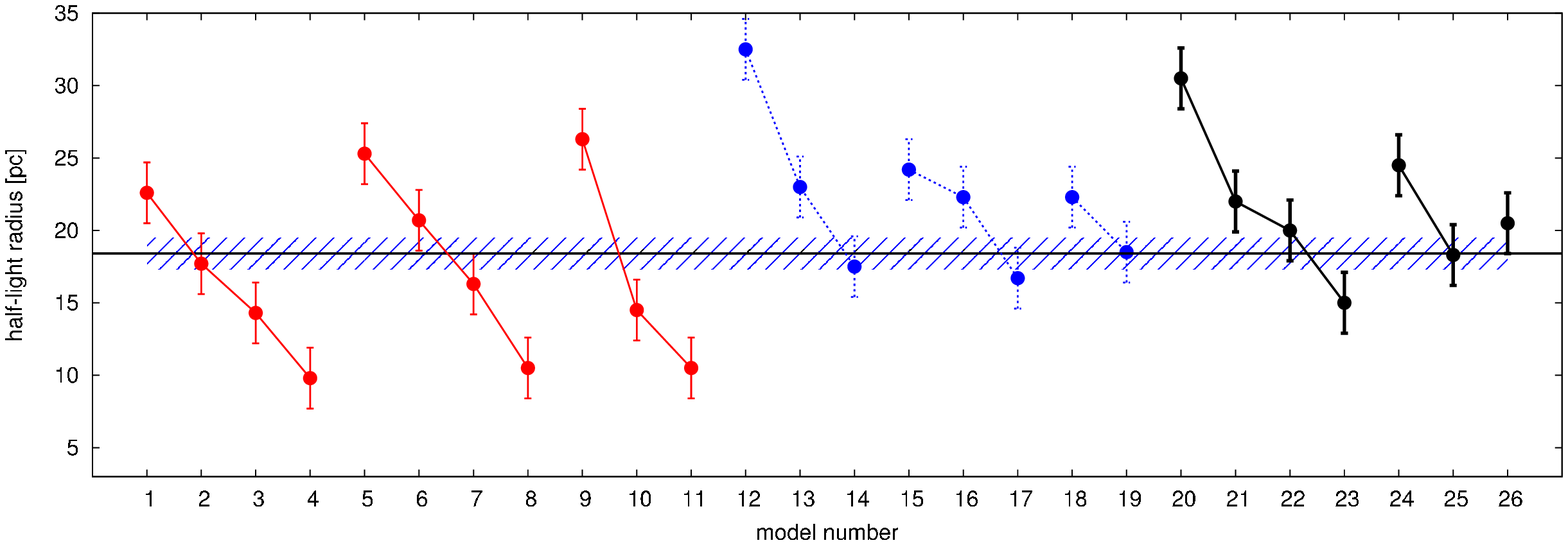}
\includegraphics[width=150 mm]{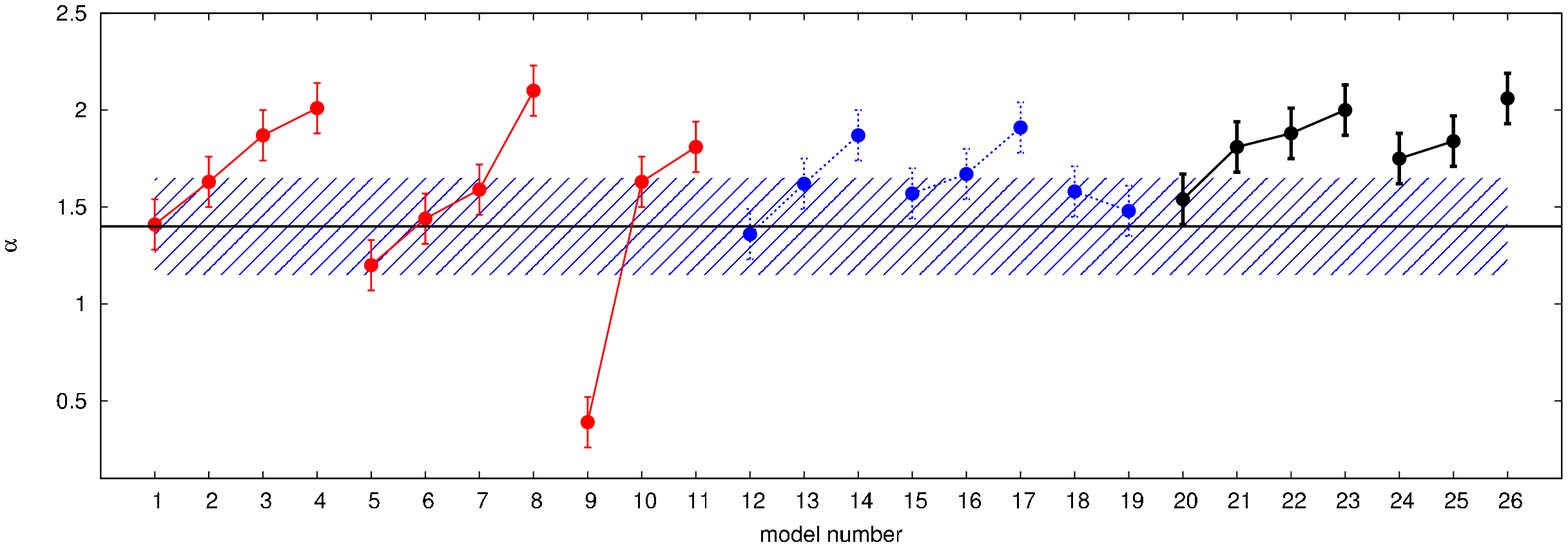}
\caption{ The total mass in the observed stellar range of $0.55<m/\msun<0.85$ (upper panel),
the present day half-light radius (middle panel), and the MF-slope (lower panel) for all models. The horizontal
axis is the model number listed in Table 2. The horizontal lines show the observed
values with the thickness of the shaded region representing the corresponding errors. The typical
numerical errors given in Table 2 are shown on the data points. Further properties
of the clusters after 11 Gyr of evolution are given in Table 2. The points
corresponding to model number 1 (marginally) and 7 are compatible with the observations within the
error bars. The error bars in the upper panel are smaller than the sizes of data points and are not shown. Different colors are associated with different pericentric distances So that}
\label{modelnumber}
\end{figure*}

\begin{figure}
\includegraphics[width=85 mm]{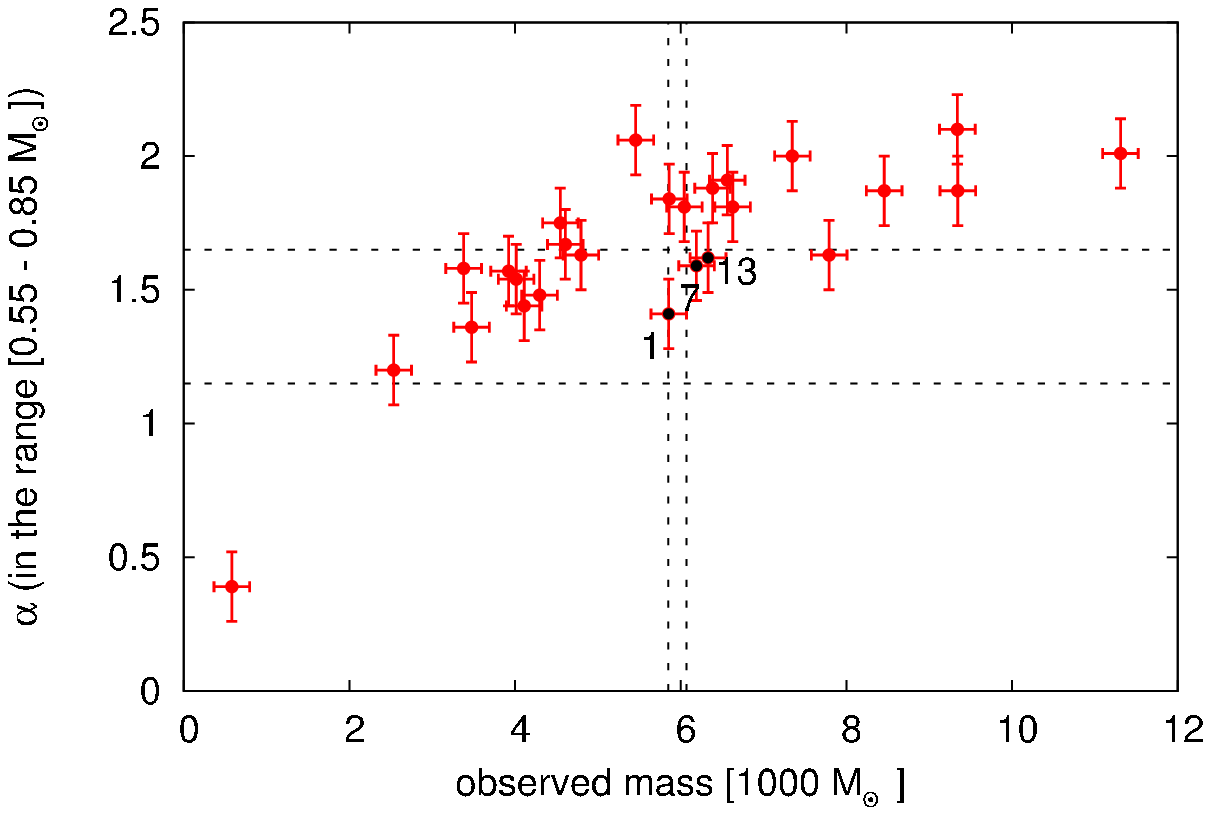}
\includegraphics[width=85 mm]{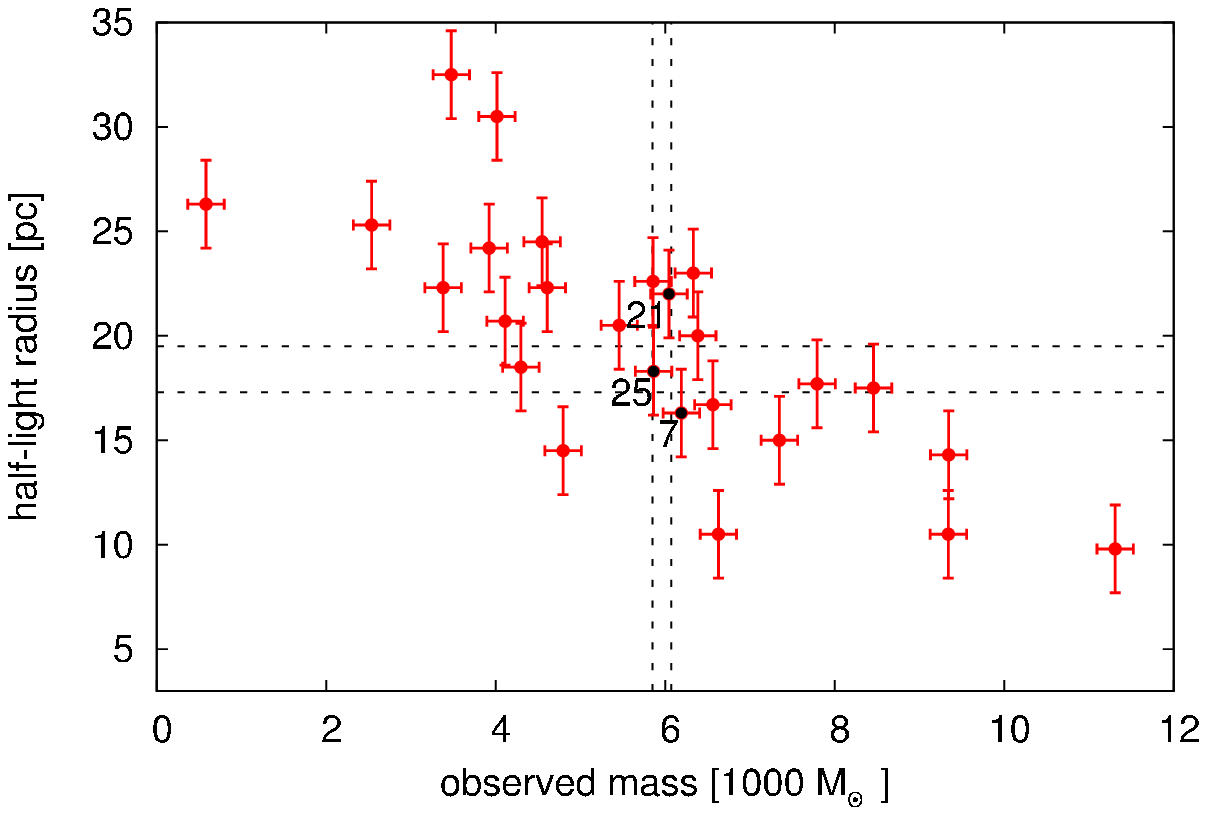}
\includegraphics[width=85 mm]{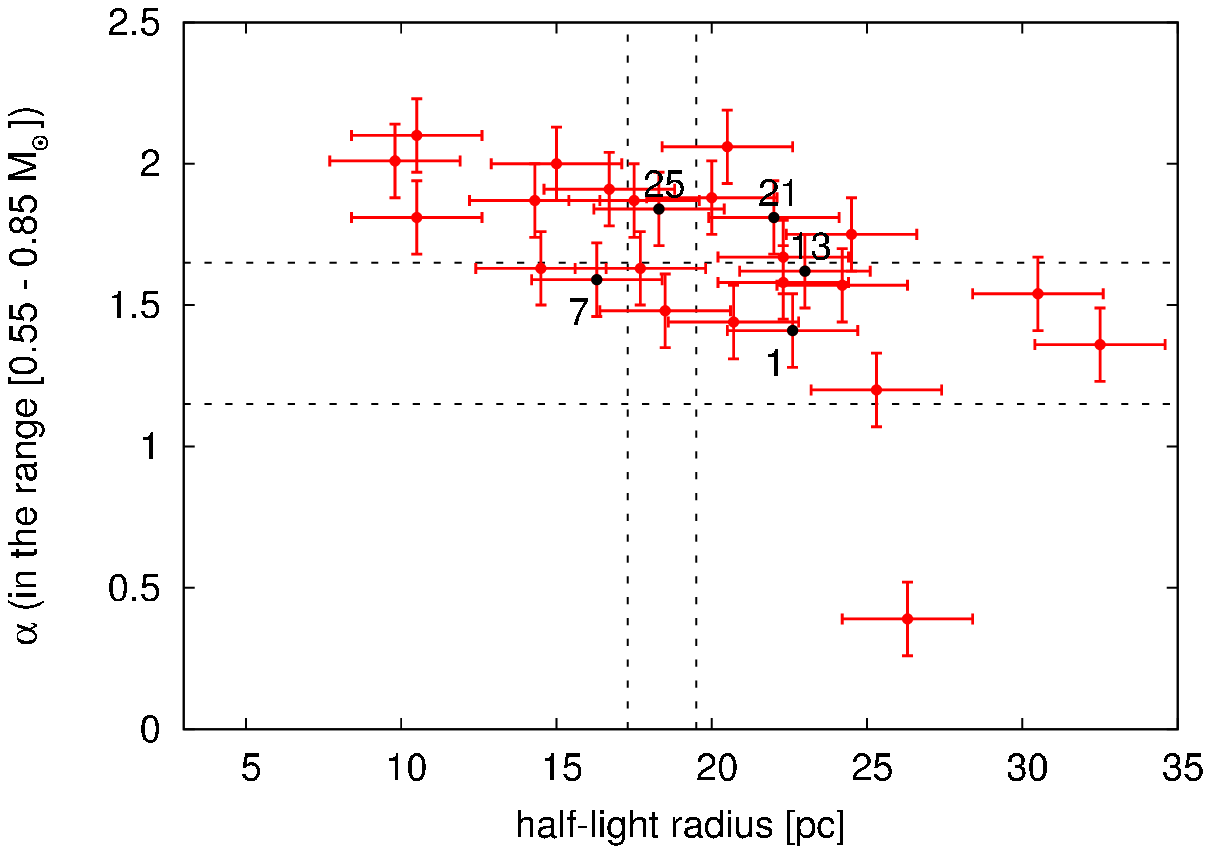}
\caption{ Plots of the total mass in the observed stellar range $0.55<m/\msun<0.85$,
the present day half-light radius, and the MF-slope for all models. The horizontal and vertical
lines show the observed values with the thickness representing the corresponding errors. The typical numerical errors given in Table 2 are shown on the data points. Additional properties
of the clusters after 11 Gyr of evolution are given in Table 2.  In each panel, only models that lie within the observed range are labeled with numbers. The point with number 7 is compatible with the observations of Pal\,4 within the uncertainties in all panels.}
\label{compare}
\end{figure}

\begin{figure}
\includegraphics[width=85mm]{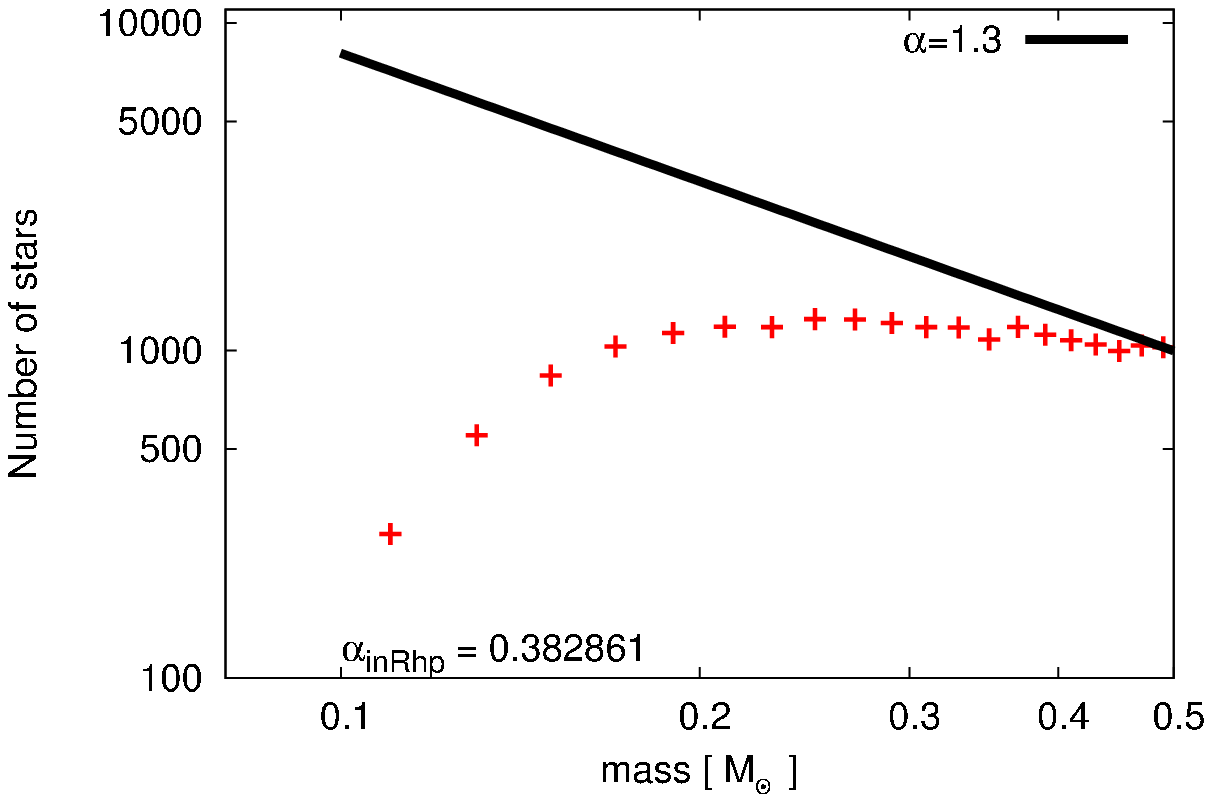}
\caption{Global MF in the mass range $M\leq0.5\msun$ for model number 7,
'M100R4', which started with $M=100000$\msun and $R_h=4$\,pc with a pericentric radius of $R_p=5$\,kpc. The mass function at the start of the simulation was chosen to be a canonical IMF, with $\alpha=1.3$ for stars less massive than 0.5\msun (black solid line). The red data points
depict the calculated present-day mass function of Pal\,4 after an evolution of 11 Gyr.} \label{mf}
\end{figure}

\section{Results}\label{Sec:Results}

In order to construct a model for Pal~4, we compare the results of N-body simulations with observational data described in Sec. 2.1.
Throughout the simulations, we keep only bound stars, and remove stars with $r\geq 2r_t$ from the computations, where $r_t$ is the instantaneous tidal radius.

The simulations start from a set of initial conditions with a number of parameters that can be varied to find the best fitting model. We have chosen the total mass, half-mass radius of the cluster, and the orbital eccentricity (by choosing the appropriate pericentric distance) as the three free parameters. Adopting different pericentric distances for the orbit of Pal~4, we evaluate whether the initially segregated cluster models can reproduce the observed  global mass function slope at the current age of Pal\,4  (i.e., 11 Gyr).

The evolution of the total bound mass and half-mass radius as a function of time is plotted in Fig. 1 for some selected models orbiting on an eccentric orbit with $R_p=5$ kpc. The horizontal lines in Fig. 1 show the present-day total bound mass, $M_{tot}$,  and 3D half-mass radius of the cluster. As can be seen the mass loss due to tidal stripping is orbit-dependent because $r_t$ is a function of the instantaneous galactocentric distance and cluster velocity. The sharp drops indicate the moment of the  cluster's perigalactic passage. During each perigalactic passage, most of the stars beyond the perigalactic tidal radius are stripped away from the cluster as a result of the shrinking tidal radius, leading to the quick decrease in bound cluster mass. As a cluster moves outward again, the number of cluster stars changes fairly smoothly.

We find that the final half-mass radii of the modelled clusters are very sensitive to their initial half-mass radii and their pericentric distances.  For example, in the case of $R_p=5$ kpc,  for models initialised with 80000\msun, the final radius changes from about 17.6 to 49.2 pc when we change the initial half-mass radius from 3 to 5 pc.

\subsection{Mass function and mass segregation}

The challenge in modelling Pal\,4 is a) reproducing the observed stellar mass function slope, which is depleted in low-mass stars and hence flatter than a canonical MF-slope, and b) the observed degree of mass segregation as a function of radius. Both are strong observational constraints on Pal\,4's past dynamical evolution.

In contrast to cluster models on circular orbits, the MF-slope decreases at a faster rate for GCs orbiting on eccentric orbits.  In fact, in a cluster that starts mass segregated, the low-mass stars are distributed in the outer regions, and hence the increased mass loss due to the eccentric orbit leads to a quicker change of the mass function \citep{Haghi14, Haghi15}. Fig. 2 depicts the MF of the best-fitting model 'M100R4' (model 7 in Table~\ref{tab_regular}), and compares it to the observed data. We ignore compact remnants for this measurement as they are too faint to be observable in Pal~4. As can be seen, the global mass function slope of this model after 11 Gyr of evolution is $\alpha=1.56\pm 0.13$, which is compatible with the observed value. This means that, in this case, the enhanced loss of low-mass stars through repeated tidal shocks is sufficient to flatten the mass function slope.

In order to compare the amount of mass segregation in our models with the observed one, we also measure the MF in dependence of the radius of the modelled
cluster. In Fig. 3 we plot the MF-slope as a function of radius in different radial bins from the center to the outer region of model 'M100R4' after 11 Gyr of evolution. We see that the mass function steepens with increasing radius, from $\alpha \approx 0.5\pm 0.23$ within a projected radius of 3 pc to
$\alpha \approx 2.1\pm 0.17$ at the largest observed radius. For comparison, the resulting surface brightness profile of the best-fitting model is shown as black crosses in Fig. 4 to compare with the observed surface brightness profile of Pal 4 (red circles). As can be seen, both surface brightness profiles agree very well.

We therefore conclude that the enhanced stripping of low-mass stars populating the outer region of the cluster can be responsible for the flattened mass function of Pal\,4 and its segregated structure, if the cluster had a canonical Kroupa IMF and evolved over 11 Gyr on an extremely eccentric orbit (e=0.9) with pericentric distance of about 5 kpc.

We plot the MF-slope for each model at 11 Gyr as a function of the fraction of remaining mass in Figure 5 to see how the mass function of each model cluster evolves. In agreement with the general findings by \cite{Baumgardt03}, and  \cite{Webb15}, the slope of the stellar mass function of the clusters  is strongly correlated with the fraction of mass that the clusters have lost as a result of two-body interactions and tidal stripping, independently of the cluster's initial mass, size and their orbit.                

In order to assess wether the models without primordial mass segregation can reproduce the observations, we finally calculate one additional simulation with $S=0$ (models 27 in Table 2). As can be seen these models keep their initial MF-slope, that is, Pal~4 in an eccentric orbit with an eccentricity of 0.9 and perigalacticon of about $5$ kpc, but without primordial mass segregation, is not able to reproduce the observed data. This is because stars with all stellar masses are equally distributed through the cluster and are equally affected by tidal stripping.  That is,  a large degree of primordial mass segregation seems really unavoidable to explain the present-day mass function of Pal~4 confirming the conclusions reached by \cite{Zonoozi11, Zonoozi14}, and  \cite{Haghi15}.

\subsection{Likely initial conditions for Pal\,4}

In order to find the most likely initial conditions for Pal 4 we tried different parameter combinations for each orbit. We have computed a grid of about 50 models with the total number of stars between 100000 and 180000, and with the different initial half-mass radius in the range $R_h=3-6.5$\,pc.  The only models which come close to the observed parameters are listed in Tab. 2.
The results of the simulated models  with primordial mass segregation for different values of the pericentric distance, $R_p$, are summarized in Table~\ref{tab_regular}.  Figure ~\ref{modelnumber} shows the present-day half-light radius, total mass and MF-slope in the observed range for all modelled clusters.  Although, different models with different initial half-mass radii and cluster masses can partially reproduce the observed data well,  a particular model with $R_h=4$ pc and $M=10^5\msun$ (model number 7 marked as boldface, 'M100R4') can reproduce all the present-day parameters of Pal\,4 better than other models. This is also shown in Fig. \ref{compare}, where the two-parameter planes are plotted.

However,  the total estimated mass ($M_{tot}$) of the best-fitting  model, 'M100R4',  is below what Frank et al. (2012) have calculated assuming that the cluster's stellar mass function is close to the canonical form at the low-mass end. Yet, the mass of our model in the observed stellar mass range (i.e., $0.55~<~m~<~0.85\msun$) is in good agreement with the observed value (see Table 1). It is therefore likely that the assumptions on Pal\,4's low-mass content made by \citet{Frank12} were too optimistic. The authors assumed that the measured mass function slope of $\alpha=1.4\pm0.25$ holds down to 0.5\msun, and have adopted a Kroupa (2001) mass function, with $\alpha=1.3$ for masses $0.08~<~m/\msun~<~0.5$, and $\alpha=0.3$ for stellar masses $0.01~<~m/\msun~<~0.08$. However, we see in our simulations that the slope of the mass function in the mass range  $m~<~0.5\msun$ is less than this value (Fig. 8). Frank et al.'s conservative mass estimate of $20100\pm600\msun$ seems to be in much closer agreement with most of our models that give an overall good fit to the observed data.

Note that the model 'M100R4' is not likely the only acceptable model and one would need to calculate a grid of models to find the best fit in the initial mass-radius plane. This is computationally too expensive with a direct $N$-body code, and also not desirable given the many simplifying assumptions we have made about the Galactic potential, Pal\,4's orbit, and the initial conditions of the cluster. Interpolating between marginally acceptable models in terms of each observed parameter we find that for models orbiting on an eccentric orbit with $R_p=5$ kpc, the best-fitted initial half-mass radius  and cluster mass are in the range $R_h=[4-5]$ pc and $M=[1-1.2]\times10^5\msun$, respectively.

Running the best-fitting model until final dissolution we find that Pal\,4 can live for another 6 Gyr. This implies that there is a significant chance to observe Pal\,4-like clusters as the MW harbors many of these extended low-mass clusters. Therefore, the eccentric orbit is a viable model to explain such clusters.

\subsection{Pal 4's proper motion}

\begin{figure}
\includegraphics[width=85 mm]{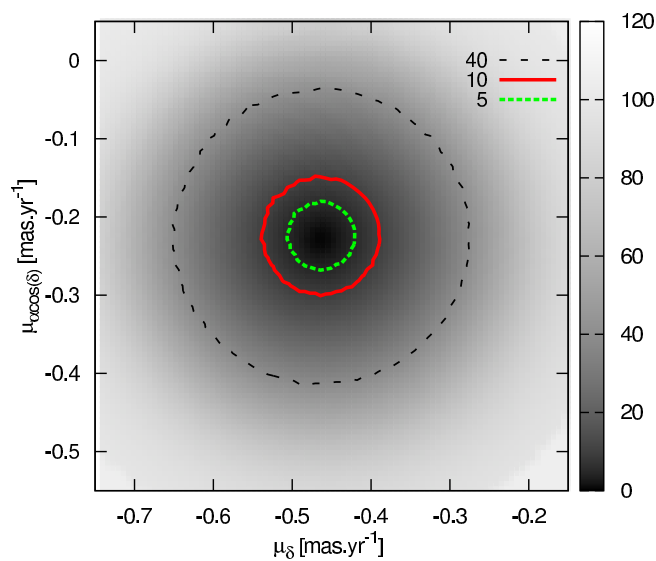}
\caption{ Perigalactic distance (in kpc) of Pal 4 versus proper motion for a line-of-sight velocity of Pal 4 of $v_{los}=72.55\pm0.22$\,km\,s$^{-1}$. For perigalactic distances less than 10 kpc, the components of the cluster's proper motion are confined to a narrow range of values (i.e., $-0.52 \leq \mu_{\delta} \leq-0.38$ mas\,yr$^{-1}$ and($-0.30 \leq \mu_{\alpha\cos\delta} \leq-0.15$ mas\,yr$^{-1}$).   }
\label{mu-PA}
\end{figure}

For a given Galactic potential, any orbit is completely specified by 6 numbers (i.e., position and velocity vectors). The proper motion of Pal 4 is unknown, hence, there could be different values of $\mu$ and $PA$ (or equivalently $\mu_{\alpha\cos\delta}$  and $\mu_\delta$) that would give similar orbits to the one we obtained. Moreover, the previous chapters have shown that the perigalactic distance of Pal 4 has to be around 5 kpc and so we check which proper motions lead
to orbits with a given perigalactic distance.
Figure \ref{mu-PA} shows that only a narrow range of $\mu_\delta$  (in the range $[-0.52,-0.38]$ mas\,yr$^{-1}$) and $\mu_{\alpha\cos\delta}$ (in the range $[-0.30,-0.15]$ mas\,yr$^{-1}$) leads to a perigalactic distance less than 10 kpc. Thus, for a given pericenter distance, $R_p$, there is a degeneracy of orbits and thus of proper motion values which can be seen in Fig. 8 as a circle (e.g. the green circle for $R_p$=5 kpc). This degeneracy comes about from a rotational symmetry about the line of sight of orbital solutions. Each orbit has the same $R_p$, passes the same focal point (the Galactic centre) and has the same line-of-sight velocity component and thus eccentricity and semi-major axis, but a different inclination.

GAIA will measure the proper motion of a handful of giant stars in Pal 4 to a precision that may suffice to test our scenario. The magnitude of the spectroscopic targets in Pal 4 fall in the range $17.5 \leq V \leq 20.0$ mag (Frank et al 2012). With the 5 year data from GAIA this will give us proper motions for Pal 4 with uncertainties of better than 0.1\,mas\,yr$^{-1}$ \citep{deBruijne14}. As such it will clearly enable distinction between a nearly circular orbit and a highly eccentric one.

\section{Conclusions}\label{Sec:Conclusions}

This paper is a follow-up to Zonoozi et al (2014) in which we modelled the dynamical
evolution of the Galactic outer-halo globular cluster Pal\,4 over its entire lifetime on a star-by star basis using the direct $N$-body code \textsc{nbody6} \citep{Aarseth03}.

In Zonoozi et al (2014) we focused on circular orbits for Pal\,4. We found that the two-body relaxation-driven loss of low-mass stars is too inefficient in such a diffuse, remote halo GC to become flattened at the high-mass end (which is  $\alpha=1.4\pm 0.25$ in the mass range 0.55-0.85\msun). We here assume an eccentric orbit for the cluster (based on its observed radial velocity and assumptions on its proper motion), to see if the enhanced mass-loss by a stronger tidal field is able to reproduce the present-day global MF-slope as well as the mass segregation of Pal\,4.

We find the best possible evolutionary model for Pal~4 by running a set of models on eccentric orbits with varying initial
half-mass radii and total masses, until an adequate fit to the observed structural parameters is obtained. The main difficulty in our calculations is to find initial models which reproduce simultaneously all these structural parameters of Pal~4.

In \citet{Zonoozi14} we concluded that, assuming a circular orbit for Pal\,4,  only models starting with a much larger half-mass radius,  with primordial mass segregation \textit{and} a flattened IMF, which is very different than what is seen for other GCs,  reach enough depletion in the global MF to be compatible with the observations.  Here (i.e. assuming an eccentric orbit for Pal\,4) we find that all observables (i.e. half-light radius, total mass, the global mass function, the degree of mass segregation and velocity dispersion) can be reproduced when starting with a canonical IMF that is primordially mass segregated. Interestingly, we have not found a scenario so far in which Pal\,4's present-day structure can be explained without primordial mass segregation. This is because the clusters expand as the result of mass loss by stellar evolution and hence the depletion of the low-mass end of the MF will speed up.

We show that eccentric orbits with $e\approx 0.9$, and $R_p\approx 5$ kpc and primordial mass-segregation cause enough expansion and enhanced mass loss of the model clusters that they match the observations. This is because star clusters lose more mass during pericentric passages on eccentric orbits, and therefore undergo stronger expansion due to some internal dynamical mechanism such as primordial mass segregation and dynamical relaxation which are dominant processes in the weaker tidal field at larger Galactocentric radii \citep{Madrid12, Haghi14} . It is therefore likely that an eccentric cluster orbit like the one we suggest has had an important influence on Pal~4's evolution. Pal\,4 had a much smaller initial size and a significantly higher birth mass than we would infer by assuming a circular cluster orbit. Our inferred initial half-light radius of about 3 to 4 pc is in the same range as the present-day half-light radii of most globular clusters, which are narrowly distributed around 3 pc \citep{Jordan05}. This is the post-gas-expulsion re-virialized clusters (Brinkmann et al. 2016) such that the birth-radius near 1 pc is consistent with the radius-mass relation of \cite{Marks10}

We also found that the present-day properties of Pal\,4 cannot be reproduced in models orbiting on eccentric orbits with $R_p\geq10$ kpc and $e\leq0.8$ (Table\,\ref{tab_regular}). Strong tidal shocks are of significant importance in our scenario.

Since Pal\,4 is a young halo globular cluster (YHGC) it most probably was formed in a tidal tail possibly drawn out from the young Milky Way in an early galaxy--galaxy encounter \citep{Zhao13, Banik16} which also spawned the disk-of-satellites (DoS, \citealt{Kroupa05, Metz09}). Together with the
satellite galaxies and tidal streams, the YHGCs form the vast polar structure around the Milky Way (VPOS, \citealt{Pawlowski11, Pawlowski13}) show that the satellite galaxies which have porper motion measurements have Galactocentric tangential velocities which are typically larger than their Galactocentric radial velocities, implying orbital eccentricities which are small. It is thus possible that Pal\,4 does not have the large orbital eccentricity implied by this work (pericenter near $5\,$kpc, apocenter $>110\,$kpc). Future proper motion measurements will clarify this issue, but if Pal 4 has an orbital eccentricity typical of the satellite galaxies then an additional mechanism would be required to flatten the stellar mass function. This mechanism is most likely gas
expulsion.  \cite{Haghi15} have shown that gas expulsion from mass-segregated infant clusters may also lead to a flattened stellar mass function. If gas expulsion is a significant physical process for these clusters then this would imply a smaller orbital eccentricity in a combination of scenarios (gas expulsion from the very young Pal 4 cluster together with an eccentric Galactocentric orbit for tidally driving additional stellar mass loss). Future numerical work will constrain the required orbital eccentricity in this scenario.

We note that Pal\,4's stellar mass function in the low mass range ($0.01\leq M/\msun \leq 0.5$) is most likely significantly depleted compared to a canonical MF (Fig. 6), and the present-day masses of these models are closer to the conservative lower limit of $20100\pm600\msun$ of \citet{Frank12}, for which the authors assumed a negative mass function slope of $\alpha_{low} = -1.0$ in this stellar mass range. Compared to the $29800\pm800\msun$ that \citet{Frank12} calculate using the canonical form of the low-mass MF, this is a significant mass correction of $\approx$50\%. Our investigation therefore shows how N-body computations can help to inform observations for regimes that are inaccessible with current telescopes.

Finally, we constrain the proper motion components of Pal 4 to be in the range $-0.52\leq\mu_\delta\leq-0.38$ mas\,yr$^{-1}$ and  $-0.30\leq\mu_{\alpha\cos\delta}\leq-0.15$ mas\,yr$^{-1}$. Since the proper motion of Pal 4 is unknown, future proper motion measurement with GAIA might be able to test our scenario.

There are some indictions that the problem that we see for Pal 4 is not limited to this cluster, and maybe all outer halo globular clusters have very depleted and segregated mass functions.  In that case, the same initial conditions are probably required for all outer halo GCs and they  would have to be on very eccentric orbits which means we could predict their orbits as well.  This could also be the case if they were accreted from a satellite galaxy.   The likelihood of this scenario would increase if these clusters are found to orbit within the VPOS (Pawlowski \& Kroupa 2014).

Recently, Bianchini et al. 2015 and Miholics et al. 2015 tried to explain the present-day radii of outer halo GCs by the accretion scenario. According to this scenario, the extended clusters form originally in the centre of dwarf galaxies which are accreted onto the MW and later expand due to a time-variation of the tidal field when the cluster escaped from the dwarf galaxy being released into the outer halo. However, it should be noted that an accreted origin of extended GCs is unlikely to explain their observed extended structure as shown by \cite{Bianchini15} and \cite{Miholics16}, supporting the idea that such clusters are already extended at the stage of their formation \citep{Haghi14}. In fact, the cluster experiences an expansion when it is removed from the compressive tides of the host dwarf galaxy but such an expansion is not enough to explain the observed extended structure, such that star clusters that underwent such a process are always less extended than the one evolved in isolation where the two-body relaxation is the dominant process \citep{Bianchini15}.  A more comprehensive assessment of this scenario in forming the flattened MF in the context of the dynamical evolution of tidally limited (N-body) models will be discussed in an upcoming study (Zonoozi et al. in preparation). 

Although, we showed that the present-day half-light radius of modeled clusters are strongly affected by the adopted initial conditions (i.e., initial half-mass radii and their pericentric distances), this is not the case for the majority of Milky Way GCs. \cite{Gieles11} found that considering circular orbits for all MW GCs, almost two-thirds of them are in the expansion-dominated phase, meaning that their evolution is not seriously affected by the Galactic tides, the eccentricity of the orbits and and the initial conditions. Indeed, the present-day mass-radius-Galactocentric distance relation of a good fraction of the MW GCs can be explained by the two-body relaxation mechanism without dependencies on the initial conditions and orbital parameters.

\section*{Acknowledgements}
The authors thank Sverre Aarseth for making his codes publicly available. AHWK acknowledges support from NASA through Hubble Fellowship grant HST-HF-51323.01-A awarded by the Space Telescope Science Institute, which is operated by the Association of Universities for Research in Astronomy, Inc., for NASA, under contract NAS 5-26555.

\bsp \label{lastpage} \end{document}